# Application of zero-noise extrapolation-based quantum error mitigation to a silicon spin qubit

Hanseo Sohn[1†], Jaewon Jung[1†], Jaemin Park[1†], Hyeongyu Jang[1], Lucas E. A. Stehouwer[2], Davide Degli Esposti[2], Giordano Scappucci[2], and Dohun Kim[1]*

[1]*NextQuantum, Department of Physics and Astronomy, and Institute of Applied Physics, Seoul National University, Seoul 08826, Korea*

[2] *QuTech and Kavli Institute of Nanoscience, Delft University of Technology, PO Box 5046, 2600 GA Delft, The Netherlands*

[†]*These authors contributed equally to this work.*

*Corresponding author: dohunkim@snu.ac.kr*

**Abstract**

As quantum computing advances towards practical applications, reducing errors remains a crucial frontier for developing near-term devices. Errors in the quantum gates and quantum state readout could result in noisy circuits, which would prevent the acquisition of the exact expectation values of the observables. Although ultimate robustness to errors is known to be achievable by quantum error correction-based fault-tolerant quantum computing, its successful implementation demands large-scale quantum processors with low average error rates that are not yet widely available. In contrast, quantum error mitigation (QEM) offers more immediate and practical techniques, which do not require extensive resources and can be readily applied to existing quantum devices to improve the accuracy of the expectation values. Here, we report the implementation of a zero-noise extrapolation-based error mitigation technique on a silicon spin qubit platform. This technique has recently been successfully demonstrated for other platforms such as superconducting qubits, trapped-ion qubits, and photonic processors. We first explore three methods for amplifying noise on a silicon spin qubit: global folding, local folding, and pulse stretching, using a standard randomized benchmarking protocol. We then apply global folding-based zero-noise extrapolation to the state tomography and achieve a state fidelity of 99.96% (98.52%), compared to the unmitigated fidelity of 75.82% (82.16%) for different preparation states. The results show that the zero-noise extrapolation technique is a versatile approach that is generally adaptable to quantum computing platforms with different noise characteristics through appropriate noise amplification methods.

**Introduction**

Fault-tolerant quantum computing requires quantum error correction to build logical qubits using large numbers of physical qubits that have error rates less than the fault-tolerance threshold. Logical qubits have been realized in superconducting qubits [1–4], trapped-ion qubits [5], neutral-atom qubits [6], and diamond-based spin qubits [7]. However, in the noisy intermediate-scale quantum (NISQ) era, the full implementation of this error correction technique is still limited in scope due to the requirement of large numbers of physical qubits with the addressability of high gate fidelity. Alternative methods such as error suppression [8–11], quantum optimal control [12–15], and error mitigation [16] have been suggested and actively utilized in the NISQ era. Especially, quantum error mitigation (QEM) is also vital in the era of fault-tolerant quantum computing due to its capability of reducing the overhead of quantum error correction [17].

Quantum dot-based spin qubit platforms [18] offer significant advantages for scalable quantum processors, including long coherence times [19] and compatibility with industrial foundry processes [20–22]. Although silicon spin qubit systems currently operate with a smaller number of qubits [22–24], recent advances in spin qubits in group-IV materials have led to demonstrations of high-fidelity operations on two-qubit gates beyond the threshold [25–27] and initial quantum error correction on three qubits [28]. These key features position silicon spin qubits as a promising platform for future large-scale quantum computing implementations. While error suppression [29–31] and optimal control [32] have been studied in spin qubits, error mitigation techniques remain less explored in spin qubit platforms despite their cross-platform viability. A recently suggested error mitigation technique termed zero-noise extrapolation (ZNE) is compelling for silicon spin qubits since it does not require any additional quantum resources and is hardware agnostic if the noise in the hardware is invariant to time rescaling [33,34]. Although the requirement is violated by the presence of time-correlated noise, such as charge fluctuations in silicon spin qubits [35–40], theoretical investigations have shown that ZNE can still effectively mitigate the incoherent errors under colored noise depending on the specific implementation conditions [41].

In this paper, we introduce the experimental application of ZNE in a silicon quantum dot qubit. We use three different methods to amplify the noise of the quantum circuits and then extrapolate the expectation values of the observable to the noise-free limits. We then examine the performances of each method using the randomized benchmarking protocol. Among the three methods, a global folding method gives accurate and stable mitigated results consistent with the previous work. Consequently, we conduct quantum state tomography (QST) along with the ZNE using a global folding method that exhibits the best performance in noise amplification with an optimized number of shots and nodes [42]. We also employ gate-set tomography (GST) to investigate the model violation and further discuss the effect of the non-Markovian noise of the model.

**Experiment**

The spin qubit device depicted in Fig. 1(a) is fabricated on an isotopically enriched (800 ppm of residual $^{29}$Si) $^{28}$Si/SiGe heterostructure [31,43,44]. The gate layers are constructed as screening gates (pink; roughly dividing regions for quantum dot array and sensors), accumulation and plunger gates (green; controlling confinement potentials), and barrier gates (blue; controlling tunneling couplings) that are electrically isolated by Al$_2$O$_3$ dielectric layers [24,45]. While the device is designed to perform five qubit control, we focus on the leftmost set of quantum dots as a qubit and a charge sensor [Fig. 1(a)]. The voltage applied to the plunger gate $V_\text{p}$ is adjusted to tune a quantum dot to the single-electron regime, thereby realizing a single spin qubit. The yellow dot represents a sensor dot (SD) capacitively coupled to a qubit quantum dot acting as a radio frequency charge sensor.

The qubit state is characterized by ground $|0\rangle \equiv |\downarrow\rangle$ and excited $|1\rangle \equiv |\uparrow\rangle$ states arising from the Zeeman splitting induced by the external magnetic field ($B_\text{ext}$). A cobalt micromagnet integrated on top of the device generates a transverse magnetic field gradient to facilitate electrical dipole spin resonance (EDSR) for coherent control of the spin state. The application of a phase-controlled microwave signal resonant with the state energy splitting makes it possible to drive the spin transitions between $|0\rangle$ and $|1\rangle$, which enables arbitrary single-qubit operations.

Fig. 1(b) illustrates a schematic of the spin readout and its result. We manipulate the energy levels through plunger gate pulsing in three stages: "initial," "operation," and "readout/load." In the "initial" stage, the Fermi level of the reservoir is set between the $|0\rangle$ and $|1\rangle$ states to ensure electron loading into the ground state energy level. During "operation," both spin levels are lowered below the Fermi level to avoid immediate tunneling after the spin resonance. A microwave pulse $V_\text{screen}$ is applied to the screening gate for spin manipulation. The "readout/load" stage employs energy-selective tunneling [46] to detect spin-flip events. We generated a Rabi chevron pattern by varying the microwave pulse frequency and duration in the operation stage, with the pattern indicating the resonance frequency of 14.6564 GHz at $B_{\text{ext}} = 439.7$ mT.

The micromagnet generates a transverse gradient needed for EDSR but also introduces a decoherence gradient that induces fluctuations in the qubit frequency. Within the condition, the device achieves 5.2 $\mu$s for the inhomogeneous coherence time $T_2^*$ and 22.3 $\mu$s for the Hahn echo coherence time $T_2^{\text{echo}}$.

**Zero-noise extrapolation**

The implementation of ZNE consists of two main steps: (1) noise amplification and (2) extrapolation. We can measure the expectation values from the quantum circuits with different noise factors by intentionally scaling the noise parameters. Then, we obtain a noise-free expectation value by extrapolating the noise-amplified values to the zero-noise limit.

The noise amplification methods can be categorized into analog and digital methods. The analog amplification technique, known as the pulse stretching method [34,47–50], is implemented at the hardware-level pulse modification that adjusts the parameters of the microwave pulse-defined gates. The technique that is commonly used for digital amplification is unitary folding, [51] which depends on the gate-level circuit composition. Considering that the state preparation and measurement (SPAM) error cannot be amplified by either the digital or analog approach, we employ the passive readout error mitigation (REM) method [28,52–55] in conjunction with the ZNE. In Fig. 2, we regard an original quantum circuit as $U$ that is composed of $N$ number of gates $G_N$ from a gate set $\mathbb{G}$ [Fig. 2(a)] and use three different methods for the noise amplifications.

The concept of pulse stretching is illustrated in the simplified schematic in Fig. 2(b), in which the microwave pulse is directly stretched from the original stretch factor $c_0$ to the desired stretch factor $c_i$. We suppose that a quantum circuit is encoded in terms of the time evolution of the time-dependent drive Hamiltonian $K(t) = \sum_\alpha J_\alpha(t) P_\alpha$ for time $t \in [0, T]$ where $P_\alpha$ is the $N$-qubit Pauli operator and $J_\alpha(t)$ is the corresponding time-dependent interaction coefficient. The expectation value from the drive Hamiltonian $K$ is defined by $E_K(\lambda)$ for the noise parameter $\lambda$. If the noise in the quantum circuit is invariant under time rescaling, the expectation value for the scaled time-dependent drive Hamiltonian $K^i(t) = \sum_\alpha J^i{}_\alpha(t) P_\alpha$ for time $c_i T$ equals an expectation value under a scaled noise $\lambda_i = c_i \lambda$, where $J^i{}_\alpha(t) = \lambda_i^{-1} J_\alpha(\lambda_i^{-1} t)$ [33]. Therefore, it is possible to obtain the noise-amplified expectation value $\hat{E}_K(\lambda_i)$ from the scaled drive Hamiltonian.

Unitary folding is a method whereby noise is scaled by inserting identity operations into the quantum circuit by exploiting the unitary nature of the ideal quantum gates. Given a unitary circuit $U$, the noise of $U$ can be amplified by that of a $(2n + 1)$-fold circuit $U(U^\dagger U)^n$, where $n$ is a positive integer. Unitary folding can be used in two ways: local folding and global folding. Local folding defines each gate inside the circuit as the given unitary circuit, as $G_1$, $G_2$, …, and $G_N$ in Fig. 2(c) so that each gate is folded for the noise amplification. Global folding considers the total quantum circuit as $U$ that is folded for the noise amplification as illustrated in Fig. 2(d).

To estimate the zero-noise limit from the noise-amplified data, Richardson extrapolation and linear extrapolation are used. Around the zero-noise value $E^*$, the expectation value of $E_K(\lambda)$ can be expanded as a power series as [33]

$$E_K(\lambda) = E^* + \sum_{k=1}^{n} a_k \lambda^k + O(\lambda^{n+1}).$$

The value of $E^*$ can be derived more accurately with Richardson's approach by reducing the higher orders of noise as

$$\hat{E}_K{}^n(\lambda) = \sum_{i=0}^{n} \gamma_i \hat{E}_K(\lambda_i)$$

where the coefficients $\gamma_i$ are defined by Lagrange polynomials as $\gamma_i = \prod_{k \neq i} \frac{c_k}{c_k - c_i}$ which have constraints of $\sum_{i=0}^{n} \gamma_i = 1$ and $\sum_{i=0}^{n} \gamma_i c_i^k = 0$ for $k = 1, 2, \ldots, n$ [33]. Moreover, according to previous theoretical work [42], the bias of the mitigated result $\hat{E}_K^n(\lambda)$ to the value of $E^*$ and the variance of $\hat{E}_K^n(\lambda)$ are closely related to the choice of noise stretch factors and the number of samplings for each factor. This relationship suggests further enhancement of the application of the Richardson extrapolation, although a tradeoff exists between bias and variance. The selection of sampling numbers and the number of nodes for the Richardson extrapolation in QST is addressed in a subsequent section. Otherwise, when the noise amplified expectation values scale linearly on the stretch factors due to the weak scaled noise, linear extrapolation can be a good substitute for the Richardson extrapolation [34,51]. Linear extrapolation is favored for the pulse-stretching method in randomized benchmarking experiments, as discussed in the following section.

**Randomized benchmarking**

We adopt a standard randomized benchmarking (SRB) protocol [56–59] to demonstrate the ability of ZNE and study the performances of the noise amplification methods. Owing to the inherent nature of SRB, it is possible to tune the noise level of the quantum circuit by setting the depths of random Clifford sequences. Furthermore, because the target noise of ZNE is an incoherent error and averaging the survival probabilities from each random Clifford sequence effectively characterizes the noise process as a depolarizing channel, SRB serves as a good testbed for the ZNE application. As the principle of ZNE holds for a time-invariant noise system, the noise amplification process is highly dependent on the noise characteristic of the hardware. In our case, $1/f$ charge noise acts as a dominant noise source in violation of this principle [35–40].

SRB is performed with ZNE based on the noise scaling methods illustrated in Fig. 2. In the case of unitary folding, the entire circuit (each gate) is folded *n* times in global (local) folding. Similar to the folding cases, in the case of pulse stretching, pulse-defined single qubit rotation gates are stretched with each stretch factor $c_i$ where $c_0 = 1$ represents the quantum circuit without the noise amplification. We choose the factors to be in the range 1 to 3 to enable us to strike a balance between the error bounds and the appropriate error scaling by considering the trade-off between them. The circuit sequences are schematically depicted at the top of Fig. 3(a)-(c). For each data point, we sample 50 random sequences, and this is repeated 1000 times for each point. To account for the limit of the finite sampling, a bootstrapping technique is used [60]. We resample each measurement result 100 times and obtain histograms of each data point which are drawn by varying the color density in Fig. 3.

For global folding and local folding [Fig. 3(a), (b)], we use Richardson extrapolation, whereas linear extrapolation is applied for pulse stretching [Fig. 3(c)] considering that the amount of scaled noise is relatively small, which makes it difficult to use the same analysis in folding cases. The results with the Richardson extrapolation demonstrate error suppression for higher-order errors as the mitigated results closely approximate the ideal expectation value. However, the bootstrapping result exhibits increasing sensitivity in the mitigated results, as the spread of each data point widens with the increasing order of extrapolation. In Fig. 3(c), within the regime of the chosen stretch factors, the linear dependence of the expectation values on the errors enables effective error mitigation through linear extrapolation.

In the presence of time-correlated noise, which is a violation of the time-invariant noise requirement in the principle of ZNE, the mitigated results in Fig. 3(a)-(c) all show improvements in terms of fidelity while the performance varies across different amplification methods. Among the three noise amplification methods, the global folding method achieves the lowest error both in the low and high error regimes. Although the local folding method delivers good performance comparable to that of the global folding method in the low error regime, its performance deteriorates drastically as the noise level increases. On the other hand, the performance of the pulse stretching method is moderate along the observed noise regime. The consistency of these findings with those of recent theoretical work [41] demonstrates that global folding effectively preserves the structure from the unfolded filter function, that is, the noise remains close to the one before time rescaling, thereby satisfying the fundamental requirements of ZNE. In contrast, the filter function scaled by the local folding method exhibits both high- and low-frequency components. The pulse-stretching method is concentrated mostly in the low-frequency regime, which violates the time-invariant assumption under noise amplification. Moreover, from the observation of stability in the mitigated results across the error regimes between the unitary folding and pulse stretching methods, it is inferred that low-frequency noise is relatively dominant compared to high-frequency noise as the circuit depth increases. This is mainly because the pulse stretching method overlaps with low-frequency noise to a greater extent than the unitary folding methods [41].

**Quantum state tomography**

The global folding method, which provides the most stable result in the presence of noise, is employed with the Richardson extrapolation method to examine the fidelities of the quantum states using ZNE. In choosing the number of nodes and the sampling numbers for the global folding method, we mainly focus on balancing the bias and the variance of the extrapolated results. The issue of exponentially increasing the variance associated with equidistant nodes [51], such as the global folding method, can be addressed by optimizing the number of samples in the Richardson extrapolation as suggested previously [42,61]. However, implementing longer sequences in our system is less favorable due to the potential accumulation of coherent errors and longer exposure to time-correlated noise. Furthermore, the sampling overhead $\Lambda$, defined by $\Lambda = \sum_i^n |\gamma_i|$ [42], which equals 2 and 3.5 for $n = 1$ and $n = 2$ in the global folding method, is relatively small to expect reduced biases for large $n$ according to [42], particularly with equidistant nodes in the non-Markovian environment. Thus, we opt for two nodes ($n = 0, 1$) for global folding with a sampling number ratio of 3:1, respectively, which corresponds to a weight ratio in the Richardson extrapolation.

We now present the experimental results obtained with the above settings. We first obtain the state fidelities without any error mitigation techniques (termed Raw in Fig. 4). Considering that the Raw data contain a SPAM error, we eliminate these errors through REM and then perform both REM and ZNE to fully improve the state fidelities. Note that in the QST process, since the quantum circuit to measure the Z component involves only one single qubit gate, its fidelity primarily depends on the calibration of the X gate. Accordingly, we do not use zero-noise extrapolation for this component and instead solely apply REM. In Fig. 4, three sets of the QST fidelities are presented for the two target states, categorized as follows: Raw, REM only, and REM and ZNE. Fig. 4(a), (b) are the results of different initial states, namely the states after the quantum gate operation of X/2 and Y/2 from $|0\rangle$ state. A schematic of the QST sequence is illustrated in Fig. 4(c). First, we prepare a $|0\rangle$ state and then execute the X/2 (Y/2) gate at the state preparation step to set the desired initial state of the tomography. The prepared state is denoted as $|-Y\rangle \equiv 1/\sqrt{2}\,(|0\rangle - i|1\rangle)$ ($|X\rangle \equiv 1/\sqrt{2}(|0\rangle + |1\rangle)$). Next, we use the I, X/2, and -Y/2 gates to perform a projective measurement along the Cartesian axes. We combine and consider the state preparation step and the tomography step as a quantum circuit $U$ which is folded for global folding in ZNE.

We repeat this sequence to obtain the state fidelities without error mitigation, with REM, and with both REM and ZNE. We take 15,000 samples for REM to obtain a spin-up probability $P_1$ to extract the readout fidelities. We can remove the readout error from the measured joint probabilities $P_M$ by applying the measured readout error correction matrix (see Appendix for details). The state fidelities are tabulated in Fig. 4(d). Without any error removal and mitigation techniques, the state fidelity is 75.82% (82.16%) for a state $|-Y\rangle$ ($|X\rangle$) due to SPAM errors and system noises (e.g. charge noises, gate noises, etc.). We primarily removed the readout error through REM and obtained an improved fidelity of 95.69% (91.76%). Lastly, we adopted ZNE into the calibrated data and achieved a 99.96% (98.52%) fidelity.

**Discussion**

The QST results show that REM significantly reduces the infidelities from 24.18% to 4.31% for the state $|-Y\rangle$ and from 17.84% to 8.25% for the state $|X\rangle$. When combined with ZNE, the infidelities are further reduced to 0.04% and 1.48%, respectively. Despite these improvements, certain limiting factors within ZNE and REM persist.

The primary constraints encountered are time-correlated and coherent noise which are noise sources that are beyond the scope of ZNE in principle. The presence of non-Markovianity within our device is detected using GST (see Appendix for details). As shown in Fig. 5, we observe a pronounced deviation from the Markovian assumptions as both the circuit depth and sequence length increase. This indicates that our system exhibits a high degree of non-Markovian noise or time-correlated noise. Systematic drifts or fluctuations in the qubit frequency due to the charge noise in the silicon spin qubit systems are one such time-correlated noise that compromises the reliability of the QEM applications [62,63]. Such instabilities in the charge noise are one of the primary factors responsible for the inconsistent performance in terms of the infidelity reduction in the QST experiments for the $|-Y\rangle$ and

|X⟩ states, and underscore the need for effective control of the charge noise to ensure reliable QEM performance. Transduced charge noises due to the presence of a micromagnet in the device can be suppressed by reducing the field gradient and real-time Bayesian estimation techniques [31,64]. Moreover, it is possible to enhance the reliability of REM by adopting the Pauli spin blockade (PSB) technique-based parity readout [24,65], which offers higher visibility and a larger readout window compared to the energy-selective tunneling technique. Despite the high visibility of the PSB readout possibly resulting in the underestimation of the requirements for REM, the practical limitations of adjusting the tunneling coupling for optimal two-qubit gate operations and PSB readout can potentially compromise the visibility. Therefore, REM remains a useful tool for enhancing the readout fidelity. Additional non-Markovian noise can be further reduced by dynamical decoupling techniques [11,66].

Coherent noise, on the other hand, can potentially be mitigated through randomized compiling (RC), a noise tailoring technique that converts arbitrary coherent noise channels into stochastic Pauli channels [67–70]. To ensure the reliability of the learned parameters in the readout error correction matrix for REM, the calibration procedure is performed immediately before the experiments to reduce the effects of readout point shift due to the charge noise.

**Conclusion**

The realization of a fault-tolerant quantum computer necessitates the implementation of quantum error correction, which imposes strenuous requirements on the number of physical qubits as well as the fidelity of quantum gate operations. Consequently, from an experimental perspective in the NISQ era, QEM can serve as a feasible approach to enhance the performance of quantum operations. In this work, we demonstrated an error-mitigating technique applicable to quantum circuits, referred to as ZNE. We first compare methods for amplifying noise processes through a standard RB protocol. For the QST experiments, we utilized the global folding technique, which exhibited optimal performance relative to the other methods evaluated. The result indicates an improvement in state fidelity to 99.96% (98.52%) compared to the unmitigated fidelity of 75.82% (82.16%) for initial state preparation $|-Y\rangle$ ($|X\rangle$). This paper presents the first experimental application of ZNE in a silicon spin qubit system and demonstrates the universal platform adaptability of the QEM technique. Furthermore, our results show the viability of achieving enhanced operational fidelities through ZNE, even in the presence of noise environments that pose challenges for precise characterization.

**Acknowledgments**

This work was supported by a National Research Foundation of Korea (NRF) grant funded by the Korean Government (Ministry of Science and ICT (MSIT)) (No. 2019M3E4A1080144, No. 2019M3E4A1080145, No. 2019R1A5A1027055, RS-2023-00283291, RS-2024-00413957, SRC Center for Quantum Coherence in Condensed Matter RS-2023-00207732, No. 2023R1A2C2005809 and Quantum Technology R&D Leading Program (Quantum Computing) RS-2024-00442994) and a core center program grant funded by the Ministry of Education (No. 2021R1A6C101B418), and Korea Institute for Advancement of Technology (KIAT) grant funded by the Korean Government (Ministry of Education) (P0025681-G02P22450002201-10054408, "Semiconductor"-Specialized University). Lucas E. A. Stehouwer and Davide Degli Esposti developed and characterized the heterostructure under the supervision of Giordano Scappucci. Correspondence and requests for materials should be addressed to DK (dohunkim@snu.ac.kr).

# Appendix

## A. Experimental setup

The sample is cooled in a dry dilution refrigerator (Oxford Instruments Triton 500) with a base temperature of ~7 mK and an electron temperature $T_e$ ~78 mK. Stable dc-voltages are applied to the gate electrodes using battery-powered dc-sources (Stanford Research Systems SIM928). The voltage pulse $V_p$ applied to the plunger gate is generated by an arbitrary waveform generator (AWG, Zurich Instruments HDAWG) with a sample rate of 2.4 GSa/s, and the vector-modulated microwave pulse $V_{screen}$ for EDSR is generated by a quantum controller Operator-X+ and Octave using the Quantum Universal Assembly (QUA) language framework by Quantum machines. Fast charge sensing is performed by rf-reflectometry [71–73] where we use a resonant LC-tank circuit connected to the ohmic contact and apply -100 dBm of carrier power at a frequency of ~130 MHz. The reflected signal is subsequently amplified by a chain of cryogenic and room temperature low-noise amplifiers and demodulated with a high-frequency lock-in amplifier (Zurich Instruments UHFLI).

## B. Readout error mitigation

For the QST experiments, we perform a calibration routine to construct a readout error correction matrix $F^{-1}$. We measure the spin-up probabilities $P_a$ and $P_b$ (a) when the qubit is initialized and (b) when the qubit is initialized, and X gate is subsequently applied to obtain the reference readout errors. We construct $F^{-1} = \begin{pmatrix} F_\downarrow & 1 - F_\uparrow \\ 1 - F_\downarrow & F_\uparrow \end{pmatrix}^{-1}$, where $F_{\downarrow(\uparrow)}$ is a spin-down (spin-up) readout fidelity. $F_\downarrow$ and $F_\uparrow$ are obtained by solving linear equations (i) $P_a = (1 - \gamma)F_\uparrow + \gamma(1 - F_\downarrow)$ and (ii) $\frac{P_b}{P_\pi} = \gamma F_\uparrow + (1 - \gamma)F_\downarrow$, where $\gamma$ is the initialization fidelity and $P_\pi$ is the expected probability of the qubit being in the spin-up state after applying the X gate to the initialized qubit. As we use an adaptive initialization [29] method, we can confirm that the charge entered the quantum dot. In addition, compared to the electron temperature $T_\text{e} \sim 78$ mK, the Zeeman splitting is equal to approximately 9 times the thermal energy; thus, we can calculate the false initialization probability to the excited state $\sim 1\%$ [74]. We calculate $P_\pi$ using the Rabi decay time and $\gamma$ using the Zeeman splitting energy and electron temperature. Finally, for the measured probability $P_\text{M}$ we obtain the probability $P_\text{C} = F^{-1} P_\text{M}$ corrected for the readout error.

### C. Gate set tomography

GST is a calibration-free [75] measurement that is self-consistent with SPAM errors. Since the gate set models of GST are based on the Markovian model, we represent the model violations as non-Markovianity and quantify the violations as goodness-of-fit according to the gate sets and the length of the gate set sequences. We illustrate the colored grid in Fig. 5, which indicates the goodness-of-fit values of the quantum circuit. The GST measurement is performed by the pyGSTi Python package using the gate sets comprising the I, X/2, and Y/2 gates.

We estimate the loglikelihood $\log L$ obtained from the measured data and compare this estimation against the maximum loglikelihood of the model $\log L_\text{max}$ to calculate the loglikelihood ratio (LLR) $2\Delta \log L \equiv 2(\log L_\text{max} - \log L)$. In the case that $2\Delta \log L$ significantly exceeds the values expected from a $\chi_k^2$ distribution of the mean $k$ and standard deviation $\sqrt{2k}$, it can be interpreted to mean that the model is violated via a non-Markovian process [75]. The gate set model assumes Markovian noise processes, thus the model violation implies the presence of non-Markovian noise dynamics. The more sensitive measurements are likely to cause higher LLR values.

In Fig. 5, when the model is violated and the LLR value of a box follows 95% of the expected $\chi^2_k$ distribution, the color of the box is depicted in linear grayscale. However, when the LLR value exceeds the range ($2\Delta \log L > 17$), the box is shown in shades of red on the logarithmic scale and we consider a significant model violation to have occurred.

**Figure captions**

**Figure 1.**

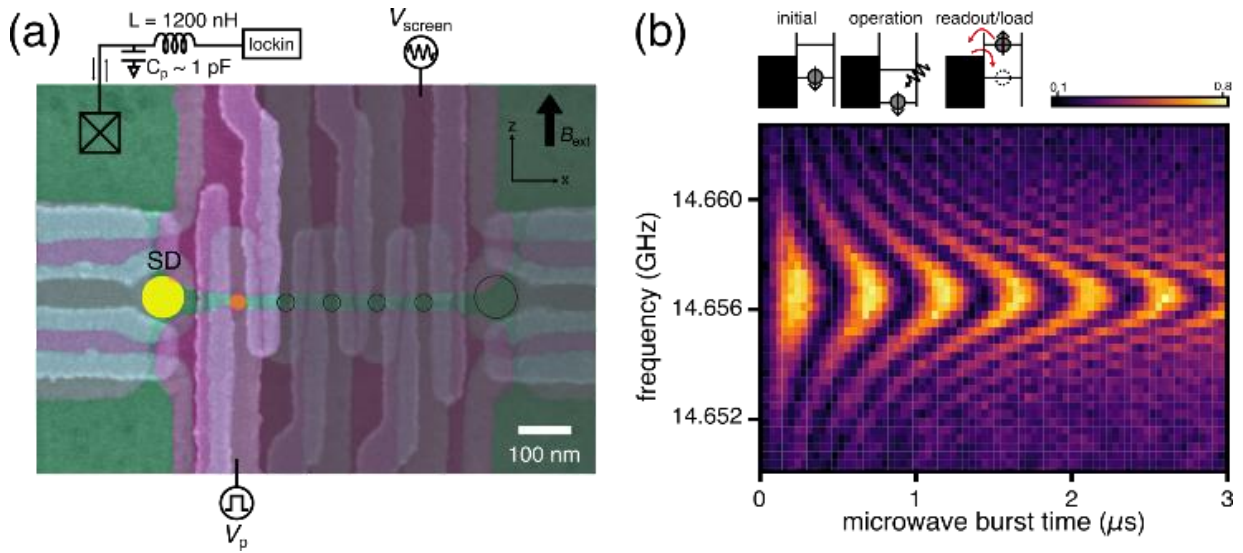

**Figure 1. The $^{28}$Si quantum dot device and spin manipulation**. (a) Scanning electron microscopy image of a semiconductor quantum dot device with a cobalt micromagnet. The device has an overlaid structure consisting of the screening gate (pink), the accumulation and plunger gates (green), and the barrier gates (blue) and is designed to form a linear five-qubit array. We focused on the leftmost dot (the small orange dot in the image) of the linear array as a qubit. By adjusting the plunger gate voltage $V_p$, we can confine a single electron in the quantum dot. The larger yellow dot indicates the sensor dot (SD) based on the radio-frequency single-electron transistor (rf-SET). An LC-tank circuit with an inductor L = 1200 nH and a parasitic capacitance $C_p \sim 1$ pF enables charge sensing via radio-frequency reflectometry. For qubit operation, we apply a microwave pulse of $V_{screen}$ to the screening gate to induce spin resonance. The thin black arrows indicate the x-z plane and the bold arrow labeled $B_{ext}$ denotes the external magnetic field. (b) Simple schematic of a single electron spin qubit operation and Rabi chevron pattern. The resonance frequency is 14.6564 GHz at $B_{ext}$ = 439.7 mT.

**Figure 2.**

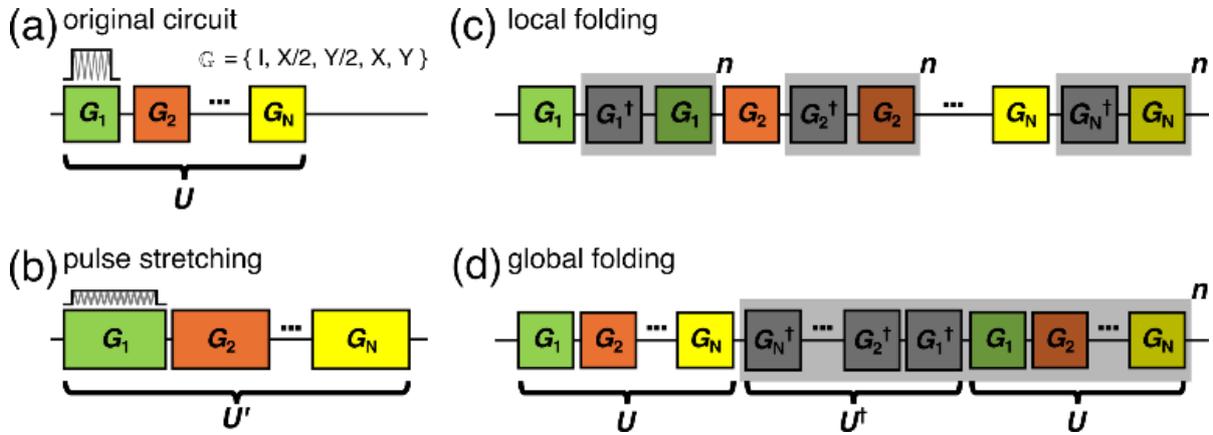

**Figure 2. Schematic of noise amplification methods.** (a) Illustration of an example of the original quantum circuit $U$, composed of the quantum gates $G_1$, $G_2$, ..., $G_N$. The square enveloped pulse shape on the gate block $G_1$ is a schematic of a microwave pulse during spin manipulation. (b), (c), and (d) represent the noise amplification methods that are used for zero-noise extrapolation (ZNE). (b) Adjustment of the microwave parameter enables the gate length of the circuit $U'$ to be extended, yet it behaves in the same way as the original quantum circuit $U$. (c) The local folding case. The gate $G_i$ is mapped to $G_i \rightarrow G_i (G_i^\dagger G_i)^n$ since $G_i^\dagger G_i$ equals the identity. (d) The global folding case, which is similar to local folding, but the entire circuit $U$ is mapped to $U \rightarrow U(U^\dagger U)^n$. In (c) and (d), the gray shaded area indicates the identity insertion part that intentionally extends the quantum circuit.

**Figure 3.**

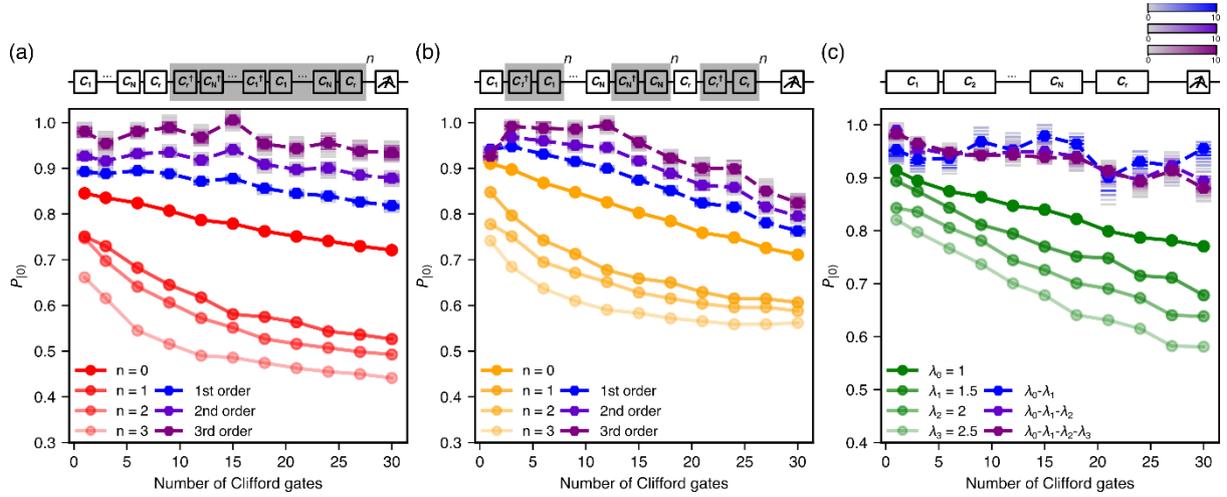

**Figure 3. Standard Clifford randomized benchmarking with ZNE.** (a)-(c) Probability of measuring a ground state with respect to identity equivalent Clifford sequences for each circuit depth. The color density plots represent the histograms of bootstrapping results of 100 samples for each data point. The noise amplification methods used are (a) global folding and (b) local folding, with corresponding $n^{\text{th}}$-order Richardson extrapolation. (c) Noise amplification using the pulse-stretching method, with the linear extrapolation results.

**Figure 4.**

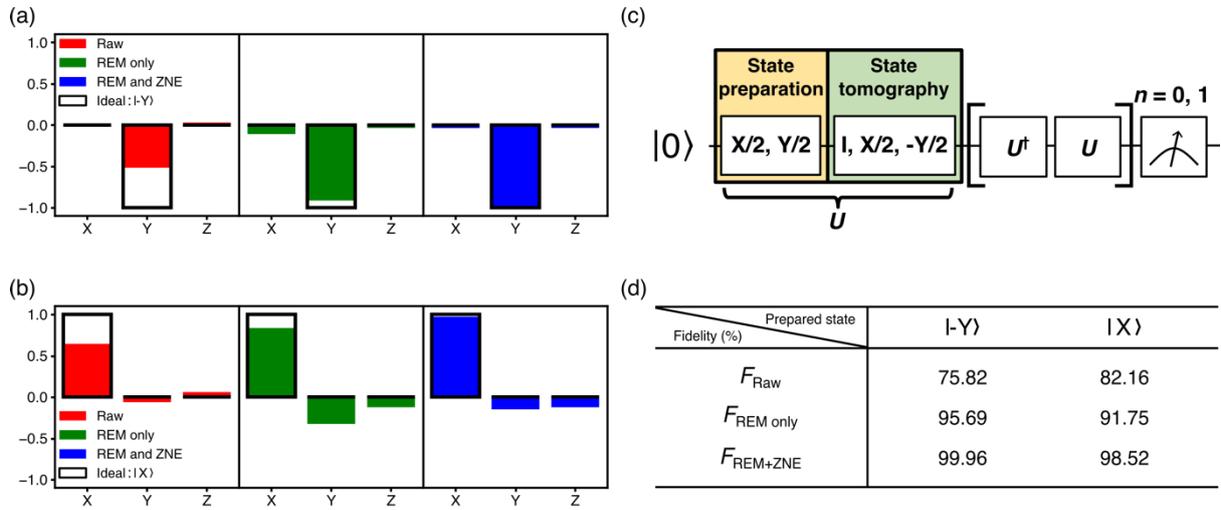

**Figure 4. Quantum state tomography result for two different prepared states: (a) $|-Y\rangle$ and (b) $|X\rangle$.** (a)-(b) Comparison of experimental and ideal values of the expectation values for each Pauli operator. The ideal values are shown as open boxes. Raw data without any mitigation, data with readout error mitigation (REM), and data with both REM and ZNE are depicted as red, green, and blue rectangles, respectively. (c) Quantum state tomography sequence for the ZNE method. The global folding method is used for the ZNE application. The number of shots for the amplification factor $n = 0$ is three times that for $n = 1$. (d) State fidelities of raw data ($F_{Raw}$), REM only ($F_{REM\ only}$), and REM and ZNE ($F_{REM+ZNE}$) with different prepared states. In both cases, the state fidelities are successfully increased through each error mitigating step.

**Figure 5.**

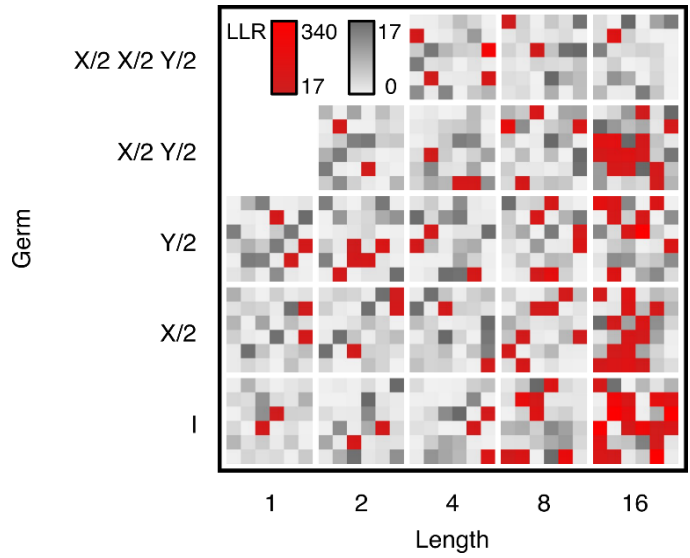

**Figure 5. Model violation box plot in gate set tomography (GST).** Each box has a loglikelihood ratio (LLR) $2\Delta \log L$ value from 36 different sets of circuits with corresponding germ circuits, where Germ refers to the short gate sets and Length to the circuit depth. If the model is Markovian, $2\Delta \log L$ is a $\chi_k^2$ random variable where $k$ is 61, 137, 254, 417, 585 for $L = 1, 2, 4, 8, 16$, respectively. The color of each box indicates the degree of model violation. Gray indicates that the model violation is within the expected values; on the other hand, red represents a significant model violation where the probability is 5% when the gate sequences are Markovian.